\documentclass[fleqn,twoside]{article}
\usepackage[headings]{espcrc2}
\usepackage{cite}
\readRCS
$Id: espcrc2.tex,v 1.2 2004/02/24 11:22:11 spepping Exp $
\usepackage{graphicx}
\usepackage[figuresright]{rotating}

\newcommand{\AmS}{{\protect\the\textfont2
  A\kern-.1667em\lower.5ex\hbox{M}\kern-.125emS}}

\title{Initial States: IR and Collinear Divergences}

\author{Martin Lavelle\address[PLY]{School of Mathematics and Statistics,
        The University of Plymouth, \\
        Plymouth, PL4 9NG, UK} and
        David McMullan\addressmark[PLY]}

\runtitle{Initial States: IR and Collinear Divergences}
\runauthor{M.~Lavelle \&\ D.~McMullan}

\begin{document}

\begin{abstract}\noindent
The standard approach to the infra-red problem is to use the
Bloch-Nordsieck trick to handle soft divergences and the
Lee-Nauenberg (LN) theorem for collinear singularities. We show that
this is inconsistent in the presence of massless initial particles.
Furthermore, we show that using the LN theorem with such initial
states introduces a non-convergent infinite series of diagrams at
any fixed order
in perturbation theory.%
\vspace{1pc}
\end{abstract}

\maketitle

\section{Introduction}

The physical origin of the infra-red (IR) problem has long been
understood: in theories with massless particles the interactions do
not fall off quickly enough. Ignoring this problem, e.g., by talking
about adiabatically switching off interactions, generates IR (soft
and collinear) divergences in S-matrix elements.

The primary theoretical responses  are twofold. For soft divergences
it is widely argued that one should use the \emph{Bloch-Nordsieck}
(BN) trick: calculate semi-inclusive cross-sections where one sums
over all \emph{emitted} soft photons with energy less than some
experimental energy resolution $\Delta$. If some charged fields are
massless (QCD or QED if the electron mass is taken to vanish) there
are also collinear divergences: then generally BN does not work and,
following the work of Kinoshita~\cite{Kinoshita:1962ur} and the
quantum mechanical theorem of Lee and Nauenberg~\cite{Lee:1964is},
it is argued that \emph{for collinear divergences} semi-inclusive
cross-sections with a sum over both initial and final particles
within some experimental angular resolution $\delta$ will be finite.

In practice it is most common to solely calculate infra-red safe
quantities, the most famous being $g-2$. If necessary the BN trick
is used while KLN with both initial and final states is very rarely
considered.

We will show that the division of just including virtual plus
emission for soft divergences but virtual, emission and absorbtion
for collinear is inconsistent. Furthermore, using the KLN theorem
for initial and final soft particles leads to an infinite series of
diagrams at any fixed order in perturbation theory. To be concrete
we consider Coulomb scattering.

\subsection{Soft and Collinear Divergences}

At one loop virtual photons produce the following soft and collinear
singularities in the S-matrix:
\begin{equation}\label{virt}
\frac1\epsilon\,,\quad\frac1\epsilon\ln(m)\,,\quad
\ln^2(m)\,,\quad\ln(m)\,.
\end{equation}
Here dimensional regularisation ($\epsilon$) regulates the soft
divergences and a small electron mass ($m$) regulates the collinear
singularities. (Full results are given in \cite{Lavelle:2005bt},
here we want to show the structure of all cancellations and
non-cancellations: hence we will only give the divergent structures
and often drop overall factors.)

The BN trick is to integrate over the emission of soft photons with
energies less than some experimental resolution $\Delta$. These soft
photons are emitted in all directions. Generally one works in the
eikonal approximation -- i.e., dropping higher powers of the photon
momentum in the numerator. The addition of this cross-section to the
one-loop cross-section which follows from (\ref{virt}) results in
the cancellation of all soft divergences $1/\epsilon\,$,
$\ln(m)/\epsilon$ and introduces a dependence on $\Delta$ into the
effective cross-section. The BN trick also cancels the leading
collinear logs, $\ln^2(m)$. However, the sub-leading collinear logs
are not cancelled and we are left, up to overall factors, with
\begin{equation}\label{virtsoft1}
-\ln(m)\times\left[\frac34-\ln\left(\frac E\Delta\right)\right]\,.
\end{equation}
Thus we have to go beyond the BN trick.

The standard argument now is that when the electron is almost
massless it cannot be distinguished from an electron accompanied by
non-soft photons emitted almost parallel to it (within an
experimental angular resolution $\delta$). Hence one should also add
semi-hard photon emission where photons parallel to the outgoing
fermion share the total energy, $E$. The diagram below thus produces
a collinear divergence
\begin{center}
\includegraphics[width=1.4cm]{bn1.mps}
\end{center}
This contributes
\begin{equation}\label{coll1}
+\frac12\ln(m)\times\left[\frac34-\ln\left(\frac
E\Delta\right)\right]\,.
\end{equation}
The KLN idea is to include absorbtion of semi-hard collinear
photons, so doubling the result of (\ref{coll1}) and  cancelling
(\ref{virtsoft1}).

This seems unnatural (including soft and semi-hard collinear
emission but only semi-hard absorbtion, i.e., no soft absorbtion),
however, there are other terms which must be taken into account. A
careful calculation yields for the contribution from semi-hard
collinear photon emission:
\begin{equation}\label{coll2}
-\frac12\ln(m)\times\left[\frac34-\ln\left(\frac
E\Delta\right)-\frac\Delta E+\frac14\frac{\Delta^2}{E^2}\right]\,,
\end{equation}
where $\Delta$ is the experimental energy resolution. What can
cancel these additional  collinear divergences? (We stress that it
is not allowed to set $\Delta$ to zero as the usual claim would then
be that the cross-section vanishes.)

It turns out that they are artifacts of the divide between soft and
semi-hard divergences. If in soft emission we go beyond the eikonal
approximation one generates additional terms which, by power
counting, must be soft finite but produce exactly these additional
collinear logs. Hence the collinear divergent contribution to the
emission cross-section is (\ref{coll1}) rather than (\ref{coll2})
but only upon inclusion of emission of those soft photon terms which
do not produce soft divergences -- the non-eikonal  $k\!\!\!/$ terms
in the intermediate fermion numerator, $p\!\!\!/+k\!\!\!/+m$ -- plus
semi-hard emission.

This is, though, still half the result needed to cancel
(\ref{virtsoft1}) and so a more precise statement of BN/KLN
cancellation for Coulomb scattering is: add all soft emission
(cancels the soft divergences and some collinear), add semi-hard
emission and semi-hard absorbtion (to cancel the collinear logs in
eq.~\ref{virtsoft1}) and also add soft absorbtion diagrams but only
retaining those terms which do not generate soft divergences and are
needed to  cancel the extra collinear logs in eq.~\ref{coll2}. This
is clearly unphysical and shows that such a divide between BN and
KLN is unacceptable.

It is thus natural to look for an approach to soft and collinear
singularities in the spirit of KLN where one includes all initial
and final degeneracies, i.e., including \emph{all} soft photon
absorbtion. As this will introduce soft divergences, we have to ask
what can remove them. For the rest of this paper we will purely
consider soft divergences and study them in the spirit of KLN.

\subsection{Emission and Absorbtion}

Since we already have all order $e^4$ contributions to the
cross-section from emission and from absorbtion, we now include all
possible diagrams with emission and absorbtion. Two of them are
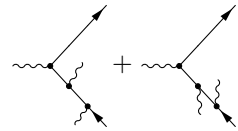
\begin{center}
\begin{tabular}{ccc}
  \begin{minipage}{.12\linewidth}
\begin{center}
\includegraphics[width=1.4cm]{disco1.mps}
\end{center}
\end{minipage} &
$+$
  \begin{minipage}{.15\linewidth}
\begin{center}
\includegraphics[width=1.4cm]{disco2.mps}
\end{center}
\end{minipage}
\end{tabular}
\end{center}
The apparent problem is that such diagrams are already at order
$e^3$. However, as noted by Lee and Nauenberg (see Appendix D
in~\cite{Lee:1964is}), there is, at the level of the cross-section a
\emph{connected interference contribution} with the following
diagram which contains a disconnected line at the level of the
S-matrix:
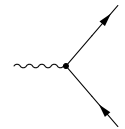
\begin{center}
\includegraphics[width=1.4cm]{disco.mps}
\end{center}
Such contributions appear rather unfamiliar but are essential for
the cancellation in Lee and Nauenberg and have been considered by a
variety of authors since their
paper~\cite{DeCalan:1972ya,Ito:1981nq,Muta:1981pe,Axelrod:1985yi,Akhoury:1997pb}.

These diagrams indeed produce a soft infra-red divergence. However,
it does not cancel the other singularities. Rather, and closely
following~\cite{Lee:1964is}, one adds either emission plus a
disconnected photon or alternatively
absorbtion plus a disconnected photon, i.e., diagrams like\\
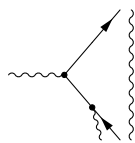
\begin{center}
\includegraphics[width=1.5cm]{abs_disco.mps}
\end{center}
When such diagrams are squared up, one obtains both connected
interference contributions
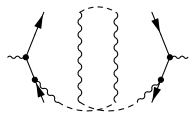
\begin{center}
\includegraphics[width=2.4cm]{disco3.mps}
\end{center}
and disconnected contributions
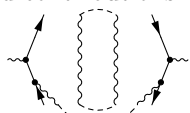
\begin{center}
\includegraphics[width=2.4cm]{disco_loop.mps}
\end{center}
If following LN one drops the disconnected terms and retains the
connected ones, the soft divergences cancel. Since this is so
important let us be explicit about how the cancellation (up to a
common factor) takes place by listing the various soft contributions
which sum to zero:
\begin{center}
\begin{tabular}{c|c|c}
  virtual & emit & absorb \\ \hline
$-\displaystyle{\frac1\epsilon}$ & $+\displaystyle{\frac1\epsilon}$&
$+\displaystyle{\frac1\epsilon}$
\end{tabular}
\end{center}

\begin{center}
\begin{tabular}{c|c}
   emit~\&\ abs. & abs.~plus
  disconn.\\ \hline
$-\displaystyle{\frac2\epsilon}$& $+\displaystyle{\frac1\epsilon}$
\end{tabular}
\end{center}
This is essentially the method used by Lee and Nauenberg and more or
less followed in~\cite{DeCalan:1972ya,Muta:1981pe,Axelrod:1985yi}.
However, is this cancellation physically meaningful? After all one
could have added either absorbtion plus a disconnected line or
emission plus a disconnected line. Indeed one could have included
\emph{more than one disconnected line at the same order in
perturbation theory}! It is thus important to study such diagrams.

\section{Many Disconnected Lines}

There are, already at this order of perturbation theory, infinitely
many connected interference contributions from the addition of
disconnected lines. Indeed only the virtual loop diagrams, with no
real emitted or absorbed photons, do not have such connected
interference contributions. Hence such interference terms need to be
taken into account in a consistent manner. To the best of our
knowledge this has been only seriously considered by
Ito~\cite{Ito:1981nq} and by Akhoury, Sotiropoulos and
Zakharov~\cite{Akhoury:1997pb} (whose proposal is essentially
identical to that of Ito).

Their idea is to include all possible disconnected photons and
combine them into connected and disconnected contributions to the
cross-section. Their claim is that the disconnected terms factor out
and that the connected terms combine in a way such that the soft
divergences cancel. To summarise their idea consider the sum of
probabilities $P_{mn}$
\begin{equation}\label{ito1}
\sum_{mn}(e+m\:\mbox{soft photons}\to e+n\:\mbox{soft photons})\,.
\end{equation}
At order $e^4$ one has to include the following probabilities
$P_{00}$ (virtual loop); $P_{01}$ (emission of a soft photon);
$P_{10}$ (absorbtion of a soft photon); $P_{11}$ (both emission and
absorbtion of a soft photon). Their assertion  is that if one sums
over all
\begin{eqnarray}\label{ito2}
\mbox{Total prob.}&=&
\sum_{mna}\frac{D(m-a,n-a)}{(m-a)!(n-a)!}\times\\ && \left[
P_{00}+P_{01}+P_{10}+(P_{11}-P_{00}) \right]\,.\nonumber
\end{eqnarray}
The terms in the square bracket are the sum of connected diagrams,
$D(m-a,n-a)$ are the factorised disconnected
terms\footnote{Disconnected at the level of cross-sections as in the
last diagram drawn above.} and the factorials are combinatorical
factors.

It is then argued that soft divergences cancel in the square bracket
and the disconnected terms will cancel by normalisation. However, it
seems very strange that the virtual terms $P_{00}$ cancel in
(\ref{ito2}) since for the virtual loop diagrams any disconnected
photons will automatically be factorised as disconnected loops in
the cross-section (there is nothing for them to connect to!)

Closer inspection shows that this is a result of writing the virtual
terms as a difference of two infinite series. These series can be
shown not to  converge. Consider  the diagram formed from
interference with disconnected photons (the last diagram but one
above). The initially disconnected lines are essentially just delta
functions and they may be \lq unravelled\rq, i.e., their
contribution is exactly equal to that of the diagram
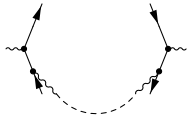
\begin{center}
\includegraphics[width=2.4cm]{appen2.mps}
\end{center}
without the disconnected line. In fact the combinatorical factors
all cancel in these connected terms. Hence in (\ref{ito2}) the
apparent factorial suppression of higher terms is indeed only
apparent.

To restate this result: for a diagram that contributes a soft
divergence of $1/\epsilon$, the same diagram with one disconnected
photon will, in the connected cross-section, also contribute
$1/\epsilon$; further every additional disconnected soft photon will
add another $1/\epsilon$. Hence we have to combine infinite,
non-converging series and the result is not well defined.

We conclude that this line of argument is not safe. Indeed, as shown
in~\cite{Lavelle:2005bt}, it can be used to argue that tree level
scattering vanishes. (At order $e^2$ there is only the probability
$P_{00}$ and in a similar fashion to (\ref{ito2}) these terms can be
argued to cancel.)

\section{Conclusions}

We have seen that in the presence of initial and final state
degeneracies it is not allowed to use the Bloch-Nordsieck trick to
handle soft divergences and the Lee-Nauenberg theorem to treat
collinear ones. This led us to consider the use of disconnected
diagrams which generate connected contributions at the level of the
cross-section. Although it is possible to produce an apparent
cancellation by only including sufficient degeneracies to do it,
such a truncation is not in any way justified. There are in fact
infinitely many contributions from disconnected photons at any fixed
order in perturbation theory. Previous attempts to sum all possible
diagrams and factor out the disconnected terms were shown to be
unsafe: the infinite series do not converge and are not well
defined.

We note that diagrams with emission and absorbtion on the same leg
produce IR finite double pole terms on that leg. It would be
interesting to see if this can be understood as a (finite) mass
shift.

It would seem attractive to consider different initial and final
states (based upon coherent states perhaps). Such work would need to
render the series convergent and cancel the soft divergences. It is
not immediately clear how this works and this is under
investigation. A completely different approach will be the subject
of D.~McMullan's talk at this conference.

We conclude that there does not exist a good understanding of the
cancellation of soft divergences (or forward collinear
divergences~\cite{Lavelle:2005bt}) in perturbation theory and that
more work in this area is urgently required.

\section*{Acknowledgments}
\noindent It is a pleasure to thank E.~Bagan for discussions on this
topic. ML also thanks the Universit\'{e} de Montpellier~II for
hospitality before the meeting and P.~Grang\'{e} for related
discussions.

\bibliographystyle{h-physrev}
\bibliography{litbank1}

\end{document}